\documentclass[prl,aps,twocolumn]{revtex4}
\usepackage{amsmath,amsfonts}
\usepackage{graphicx}
\usepackage{epstopdf}
\usepackage{color}
\usepackage{bm}

\begin{document}

\title {Orbital ordering induced ferroelectricity in SrCrO$_3$}
\author{Kapil Gupta}
\affiliation{ S. N. Bose National Centre for Basic Sciences, Block- JD, Sector-III, Saltlake, Kolkata- 700 098, India}
\author{Priya Mahadevan}
\email {priya@bose.res.in}
\affiliation{ S. N. Bose National Centre for Basic Sciences, Block- JD, Sector-III, Saltlake, Kolkata- 700 098, India}
\author{Phivos Mavropoulos}
\affiliation{Peter Gr\"{u}nberg Institut and Institute for Advanced Simulation, Forschungszentrum J\"{u}lich and JARA, D-52425 J\"{u}lich, Germany}
\author{Marjana Le\v{z}ai\'c}
\affiliation{Peter Gr\"{u}nberg Institut and Institute for Advanced Simulation, Forschungszentrum J\"{u}lich and JARA, D-52425 J\"{u}lich, Germany}

\begin{abstract}
Using density functional theory calculations, ultrathin films of SrVO$_3$(d$^1$) and SrCrO$_3$(d$^2$) on SrTiO$_3$ substrates 
have been studied  
as possible multiferroics. Although both are metallic in the bulk limit, 
they are found to be insulating  as a result of orbital ordering 
driven by lattice distortions at the ultrathin limit. 
While the distortions in SrVO$_3$
have a first-order Jahn-Teller origin, those in SrCrO$_3$ are ferroelectric in nature. This
route to ferroelectricity (FE) results in polarizations comparable with 
conventional ferroelectrics.
\end{abstract}


\maketitle
Recent advances in the growth of oxide thin films have allowed one to engineer
both electronic and magnetic functionalities in oxides, very different from 
their bulk counterparts\cite{oxidefilms,srruo3,priya,fujimori,lanio3}. This additional dimension of oxide research has
opened up the possibility of generating cross-correlated electronic couplings
which could be interchangeably controlled by both electric as well as magnetic
fields, generating new classes of multiferroics. In contrast, there are very 
few examples among bulk materials which are both magnetic as
well as ferroelectric. This is because FE is usually found in d$^0$
systems and d$^0$-ness is not compatible with magnetic order \cite{nicolarev}. There are examples
of systems which exhibit both orders among finite d$^n$ systems, however the 
magnitudes of polarization are usually small in such cases. 

An avenue that has been quite successful as a route to new ferroelectrics 
in thin films is one where substrate strain and overlayer
thickness have been used as handles to modify properties resulting in behavior 
very different from the bulk. The work by Haeni {\it et al.}~\cite{haeni} showed that
substrate strain could be used to engineer a ferroelectric ground
state in SrTiO$_3$, a $d^0$ paraelectric. While it was not obvious that this route 
should work for multiferroics also, first-principles calculations~\cite{Lezaic2011} 
predict that the strain induced by merely depositing the room-temperature 
ferrimagnetic double perovskites Bi$_2$NiReO$_6$ and 
Bi$_2$MnReO$_6$ on a substrate should turn their antipolar bulk ground state into 
a ferroelectric one. 
S. Bhattacharjee~{\it et al.}~ \cite{ghosez} showed theoretically that by increasing the lattice
constant of CaMnO$_3$, a polar mode becomes soft 
and almost degenerate with an antipolar mode. 
Similar observations were made in epitaxially strained
simple binary oxides such as BaO, EuO~\cite{bao} as well as BaMnO$_3$~\cite{james},  
SrMnO$_3$~\cite{rabe} and EuTiO$_3$~\cite{Fennie,Lee}, none of which are examples of  bulk  ferroelectric materials.
All these examples were band insulators. Therefore, Jahn-Teller effects
were irrelevant in these systems, allowing, under strain, 
for the manifestation of the 
pseudo Jahn-Teller effects, which are usually associated
with non-centrosymmetric distortions. An important consequence
of these studies was that ferroelectric polarization as large 
as that found in $d^0$ ferroelectrics was predicted in these
oxides. In contrast the observed ferroelectric
polarization in most multiferroics  is at least one or more orders of
magnitude lower\cite{pol1,pol2,pol3}. This route to a high polarization  and possibly 
magnetism associated with the same atom opens up interesting possibilities. 

Orbital ordering has recently emerged as an alternate route 
to FE. 
This was recently realized in films of doped manganites on a high-index surface of 
the substrate~\cite{miyano} where orbital ordering with an axis tilted away 
from the substrate normal was found. This led to off-centring. Again in this case the
examples considered were bulk insulators, which we show in the present work
need not be a limitation in the choice of materials. We start by considering the
example of SrVO$_3$ and SrCrO$_3$, both of which are metallic in the bulk. In this work
we explore a new route to non-$d^{0}$ 
FE through ultrathin films of transition metal oxides
which, as discussed earlier, were found to be insulating. In contrast to the examples listed above,
in the cases we investigate, the transition-metal cation is Jahn-Teller active in bulk. 
A further important difference is that strain (although it could be present) does not play a crucial role in 
the induced FE.

Ultrathin films (just two monolayers(MLs)) 
of SrVO$_{3}$ on SrTiO$_{3}$~(001) as well as films comprising of two MLs 
of SrCrO$_3$ on SrTiO$_3$~(001) are considered.
Examining the electronic structure theoretically, we find both systems to be insulating,
both arising from an orbital ordering at the surface, driven primarily by structural
distortions.
While SrVO$_3$ favors Jahn-Teller distortions, SrCrO$_3$ favors a
ferroelectric ground state, with a large contribution to the energy lowering in the insulating 
state derived from these distortions. 
Our detailed analysis shows that the modifications of the crystal field at the surface associated
with a missing apical oxygen allow for the observed FE in SrCrO$_3$ (and possibly in other 
non-$d^{0}$ transition metal compounds). These distortions at the surface survive even for the thicker films.
The calculated polarization is found to be 
large, comparable to that  in bulk $d^{0}$ ferroelectrics. 

We consider a symmetric slab (see for instance Ref. ~\cite{vanderbilt}) 
consisting of 
13 layers of SrO - TiO$_{2}$, terminating in a TiO$_2$ layer, 
to mimic the substrate. 
The in-plane lattice constant is fixed at the 
experimental value (3.905 \AA) for SrTiO$_{3}$, while the 
out of plane lattice constant was varied.
We  allowed for tilts of the TiO$_{6}$ as well as $M$O$_{6}$ octahedra, a distortion observed in low temperature SrTiO$_{3}$~\cite{tilt}. 
The electronic structure of these systems was calculated 
within VASP~\cite{vasp} using a k-point mesh of 6x6x1
using GGA\cite{gga}+U scheme \cite{dudarev} 
with a U$_{eff}$ of 2.2/2.5 on V~\cite{ferdi}/Cr. A 
vacuum of 15 \AA\ was used to minimize interaction between 
slabs in the periodic supercell geometry. 
Internal positions were allowed to relax to their minimum energy value. 
We construct maximally localized Wannier functions (MLWF) and use the shifts of their centers in the ferroelectric phase with respect to the paraelectric one to calculate the ferroelectric polarization~\cite{king-smith}.  For this purpose, the electronic ground state 
for the relaxed structure was converged with the Full-Potential 
Linearized Augmented Planewave Method-based code \texttt{FLEUR}~\cite{Fleur}, 
in film geometry, on a 7x7 k-points mesh. MLWF were constructed with the \texttt{Wannier90} 
code~\cite{mostofi} and the interface between \texttt{FLEUR} and
\texttt{Wannier90}~\cite{freimuth}. The advantage of the Wannier functions approach (over the Berry phase one) in the specific system where the ferroelectric polarization is not uniform is that the contributions of every layer/atom (see \cite{freimuth}) to the electronic polarization can be extracted. 

We started by calculating the electronic structure of bulk 
SrVO$_{3}$ and SrCrO$_{3}$ within GGA+U. Both compounds were
found to be metallic, consistent with experiment. 
Next, we performed similar calculations for two MLs
of Sr$M$O$_{3}$ ($M$= V, Cr) on SrTiO$_{3}$, and the ground state was in both 
cases found to be insulating. 

Let us first examine in more detail the case of  SrVO$_{3}$. 
Each VO$_5$ unit at the surface has in-plane 
V-O bondlengths (1.95 \AA) longer than the out of plane one (1.91 \AA). This could in principle 
result in a lower energy for the $d_{xy}$ orbital than for the 
doubly degenerate 
$d_{yz}$ and $d_{xz}$ orbitals (Fig. 1(a)).
As there is just one electron associated with the V$^{4+}$ ion, in this case an
insulating ground state is expected.
However, since one of  the apical oxygens is missing at the surface, 
this could result in a reversed ordering of the levels 
leading to a metallic state (Fig. 1(b)). 
Our calculations find the scenario
discussed in Fig. 1(b) realized. However, on 
performing a complete structural optimization 
one finds an orbital ordering transition which has a Jahn-Teller origin
with two elongated V-O (2.18 \AA) and two contracted V-O 
bondlengths (1.78 \AA) (see Fig. 1(c)). A ferromagnetic state is found to 
be favored by 60 meV/V at the two ML limit in contrast to the bulk which is 
paramagnetic.

To contrast the results for SrVO$_3$ films, we have also
considered two MLs of SrCrO$_3$ on SrTiO$_3$ substrate.
Here also the degeneracy lifting of the Cr $t_{2g}$ levels is a consequence of the missing apical oxygen 
with the $d_{xz}$ and $d_{yz}$ levels at lower energy than the
$d_{xy}$ level. As there are two electrons in the $d$ levels of Cr$^{4+}$ ion, they occupy the doubly degenerate $d_{xz}$ and $d_{yz}$ levels, leading to a band insulator.
Interestingly, we find that carrying out complete structural 
optimization results in a strong modification of the in-plane Cr-O bondlengths from their starting values:
while two of the bondlengths contract, the other two
extend (Fig. 1(d)) \cite{Suppl}. 
The effect of this structural 
distortion is that the system has no
inversion symmetry now and possesses a finite electric polarization. 
While in both SrVO$_3$ and SrCrO$_3$ the dominant energy lowering comes 
from the in-plane distortions, note the important difference: in 
SrCrO$_3$, the Jahn-Teller effect is quenched due to one additional 
electron on Cr$^{4+}$ cation and to the modified crystal field at 
the surface (with respect to that of the bulk material), so the 
pseudo-Jahn-Teller effects become operative driving the system 
ferroelectric/polar.
These results put ultrathin films of SrCrO$_{3}$ in the same category of band insulators turned 
ferroelectrics where orbital ordering drives the system insulating.
 Now every band insulator does not turn ferroelectric,
so this immediately brings up the question of the microscopic interactions that are
reponsible for FE here. 
In the following, we investigate the origin of the observed ferroelectric 
distortions.

In Fig. 2(a) we have plotted the partial density of Cr-$d$ states
for the structure in which the film is constrained to be 
centrosymmetric. A band gap barely opens
up in this structure between the majority spin Cr $d_{yz}$/$d_{xz}$ states and the $d_{xy}$ states.
The bandwidths associated with the $d_{xz}$, $d_{yz}$ orbitals are a factor of two less than the
bandwidths associated with the $d_{xy}$ orbitals 
as a result of broken periodicity along the positive $z$-direction.
 Allowing for the constraint of
centrosymmetry to be relaxed, we find that two of the in-plane Cr-O bondlengths become shorter
and are equal to 1.78 \AA while two others become longer and are equal to 2.15 \AA, 
in contrast to their undistorted values of 1.95 \AA. A part of the energy lowering in the 
process comes from the increased hopping interaction due to shorter Cr-O bonds. This is clearly
seen in the increased separation of the bonding and antibonding states with dominant contribution
from the $d_{xy}$ orbitals (Fig. 2(b)). The shorter Cr-O bondlengths also result in an
increased repulsion between the electrons on Cr and O. This is partly overcome by the presence of the
vacuum in the z-direction which allows the $d_{xz}$, $d_{yz}$ orbitals to extend into the vacuum
region. This may be seen from the charge density plot associated with the antibonding 
states with $d_{xz}$/$d_{yz}$ character plotted in the inset of Fig. 2(b). The lobe of the $d$
orbital closer to the longer Cr-O bond, extends further into the vacuum to compensate for the
increased repulsion associated with the shorter Cr-O bonds. These results also provide us
with design principles for generating multiferroic materials in which 
one has $MO_5$ structures resulting from MO$_6$ octahedra.
In-plane distortions as well as larger spacings in one of the $z$-direction 
in which the oxygen is absent could result in ferroic distortions.

In Fig. 3 we show the calculated values of the layer-resolved ferroelectric polarization for the case of two MLs of SrCrO$_3$ on SrTiO$_3$. The bars with vertical, (violet) pattern show the total (ionic plus electronic) 
polarization in CrO$_2$/TiO$_2$ layers, while the ones with the horizontal (green)
pattern show the total polarization in SrO layers. The (gray) bars 
outlined with violet (CrO$_2$/TiO$_2$ layers) or (green) (SrO layers) indicate 
the electronic contribution to the total polarization in the respective 
layers. Note that the electronic contribution in CrO$_2$/TiO$_2$
layers is rather large. It constitutes about 50\% of the total polarization, while in SrO layers it is smaller and negative.
The polarization points in the \textless110\textgreater ~direction (in-plane) and is calculated per layer, i.e.  
instead of the unit cell volume that would be used to evaluate it in a  
bulk system, we use the volume of each layer, calculated as the product of the area  
of the in-plane unit cell and the sum of the half-distances to the neighboring  
layers on either side. The volume of the surface layer is taken to be  
equal to the one beneath it. 
While not ferroelectric in bulk, we find that a significant polarization is induced in high-k SrTiO$_3$ by the 
ferroelectric shifts of Cr$^{4+}$ ions at the surface. Note that the polarization in the center of the film is 
constrained to zero by the chosen geometry (in the calculation, the Cr$^{4+}$ ions move in the opposite 
directions on the opposite sides of the film) and its decay with the distance from the surface need not be as rapid 
in the case of a thicker substrate. A rough estimate of the energy for switching is
found to be 42 meV \cite{details}, similar to other displacive ferroelectrics \cite{cohen_nature}.

In SrCrO$_3$ the observed ferroelectric distortions were associated with the 
two MLs  
limit. However when an additional SrO ML was added, this led to the out-of-plane
distortions becoming the largest distortions. The differing environments on either side of
the surface CrO$_6$ octahedra lead to unequal bondlengths of 1.88 \AA\ and 1.95 \AA\ along the positive and negative
z-direction. As a result, a finite dipole moment develops (though it is unswitchable by an electric field).
Similar observations have been made on other interfaces~\cite{ramesh}.
The system still remains insulating at this limit. Adding another layer we find that the surface
CrO$_2$ layer sustains ferroelectric distortions in addition to charge ordering. Cr in CrO$_2$ is found to have a negative
charge transfer energy \cite{sawatzky}. Here also we find this to be the 
case, and consequently some of the holes are localized on the 
oxygens. 
The surface Cr atoms are found to have a valency of $d^{1+\delta}$ and $d^2$ 
(Fig. 4 (a)-(b)), while the sub-surface Cr atoms have a breathing 
mode type distortion 
which leads to a valency of $d^{3-\delta}$ and $d^2$ on these Cr 
atoms (Fig. 4 (c)-(d)).
By comparing the total energy of several configurations we find a 
ferromagnetic coupling
parameter in the surface Cr layer and an antiferromagnetic coupling 
in the sub-surface layer and between the layers.

In the preceding discussion
the substrate imposed a tensile strain on the SrCrO$_3$ overlayers. It 
is impressive, however, that one finds that as a consequence
of the large stabilization energy for the in-plane ferroelectric distortions, the films can sustain these 
distortions even for 2\% compressive strain.

Having established FE at the ultrathin limit, we went on to examine the magnetic
properties as a function of thickness. The bulk is found to be an antiferromagnetic metal, consistent
with experiment. At the ultrathin limit of two MLs the antiferromagnetic state is more stable
than the ferromagnetic state by 45~meV/Cr atom corresponding to a 
nearest-neighbor coupling of $J=11.25$ meV in a Heisenberg model.
The stabilization of an antiferromagnetic state can be easily understood 
within a superexchange picture as there are channels for delocalization 
only for an antiferromagnetic arrangement of spins. This difference is 
enhanced at the three MLs limit to 
52~meV/Cr atom. At the four MLs limit various magnetic
configurations were considered, and we find that
the surface layer becomes ferromagnetic while the subsurface
CrO2 layer becomes antiferromagnetic with the
system found to be insulating. The closely competing
configuration where both surface layer Cr atoms as well
as the sub-surface layer Cr atoms are coupled antiferromagnetically
is found to be 9 meV/Cr atom higher.

From a device perspective, the important question is whether there 
is a coupling between the electrical and 
magnetic degrees of freedom. We compute the magnetic crystalline anisotropy for the two 
MLs film of SrCrO$_{3}$ and find that the spin favors an in-plane orientation over an 
out-of-plane one by 128 $\mu$eV per Cr atom. Within the plane, the spin favors an orientation at 
90$^\circ$ to the direction of the electric polarization by 14 $\mu$eV. 
We note that a similar arrangement was observed in BiFeO$_{3}$\cite{ramesh1}. 
By Monte Carlo simulations on a Heisenberg model basis
including single-ion anisotropy we find that the coupling 
between electric polarization and staggered-magnetization
direction persists up to a temperature of 70~K after which short-range magnetic order is lost. Even if this temperature
is considerably lower than room temperature, the novel mechanism that is revealed here could emerge in other
materials at higher temperature provided that the exchange is stronger.

 Ultrathin films of d$^1$ and d$^2$ transition metal oxides have been explored as possible
candidates for multiferroicity. Orbital ordering allows one to realize an insulating surface
even in systems where the bulk form is metallic. 
While these have a Jahn-Teller origin in the case of SrVO$_3$,
 pseudo Jahn-Teller effects result in sizeable ferroelectric distortions for SrCrO$_3$ 
which, even in the absence of strain,
also turns out to be an antiferromagnetic insulator at this limit of two MLs. These
distortions persist into the substrate and survive for the surface CrO$_5$ layer even for thicker films of four MLs
opening up new possibilities for non $d^0$ multiferroics.

We acknowledge useful discussions with D.~D. Sarma, S. Bl\"ugel and I.~B. Bersuker.
KG acknowledges CSIR, India and PM acknowledges DST, Government of India through the Indo-EU project ATHENA for financial support.
ML gratefully acknowledges the financial support of the Young Investigators
Group Programme of the  Helmholtz Association, contract VH-NG-409, the support of the
J\"{u}lich Supercomputing Centre and the  European Community's Seventh Framework Programme
(FP7/2007-2013) under Grant Agreement No. NMP3-LA-2010-246102.
PM and ML also thank the DST-DAAD programme which funded the exchange visits.


\renewcommand
\newpage

\begin{figure}
\includegraphics[scale=0.5,keepaspectratio=true,angle=0]{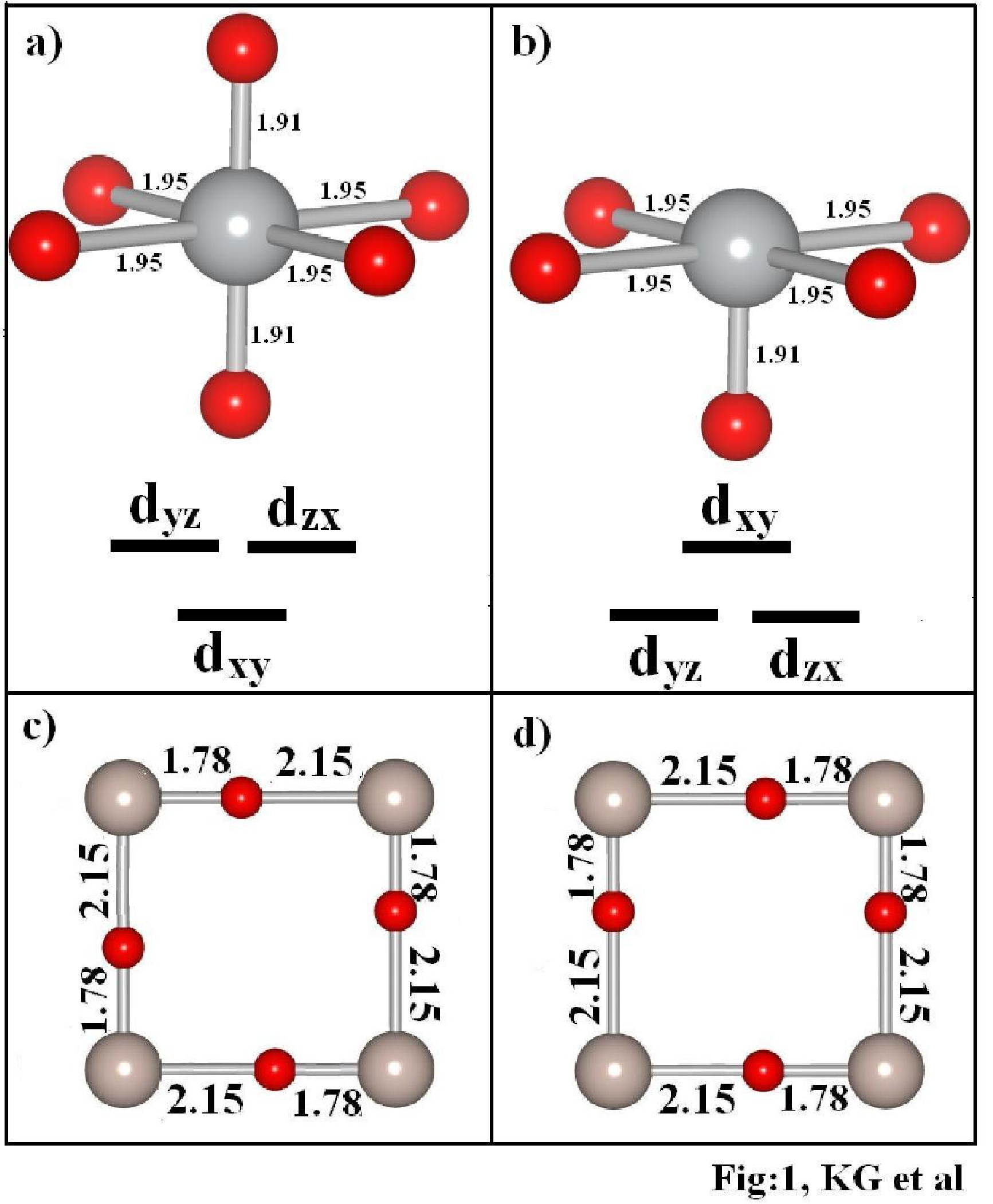}
\caption{(Color online) Distortions and crystal field splitting of the t$_2g$ orbitals for
 (a) a $M$O$_6$ octahedron under tensile strain, (b) a surface $M$O$_5$ unit, 
where $M$=transition metal atom.
The $M$-O network in the x-y plane found in the presence of (c) 
Jahn-Teller distortions (seen in SrVO$_{3}$), (d)
pseudo Jahn-Teller distortions (seen in SrCrO$_{3}$). 
Large(small) and grey(red) spheres denote M (O) atoms.}
\end{figure}

\begin{figure}
\includegraphics[scale=0.5,keepaspectratio=true,angle=0]{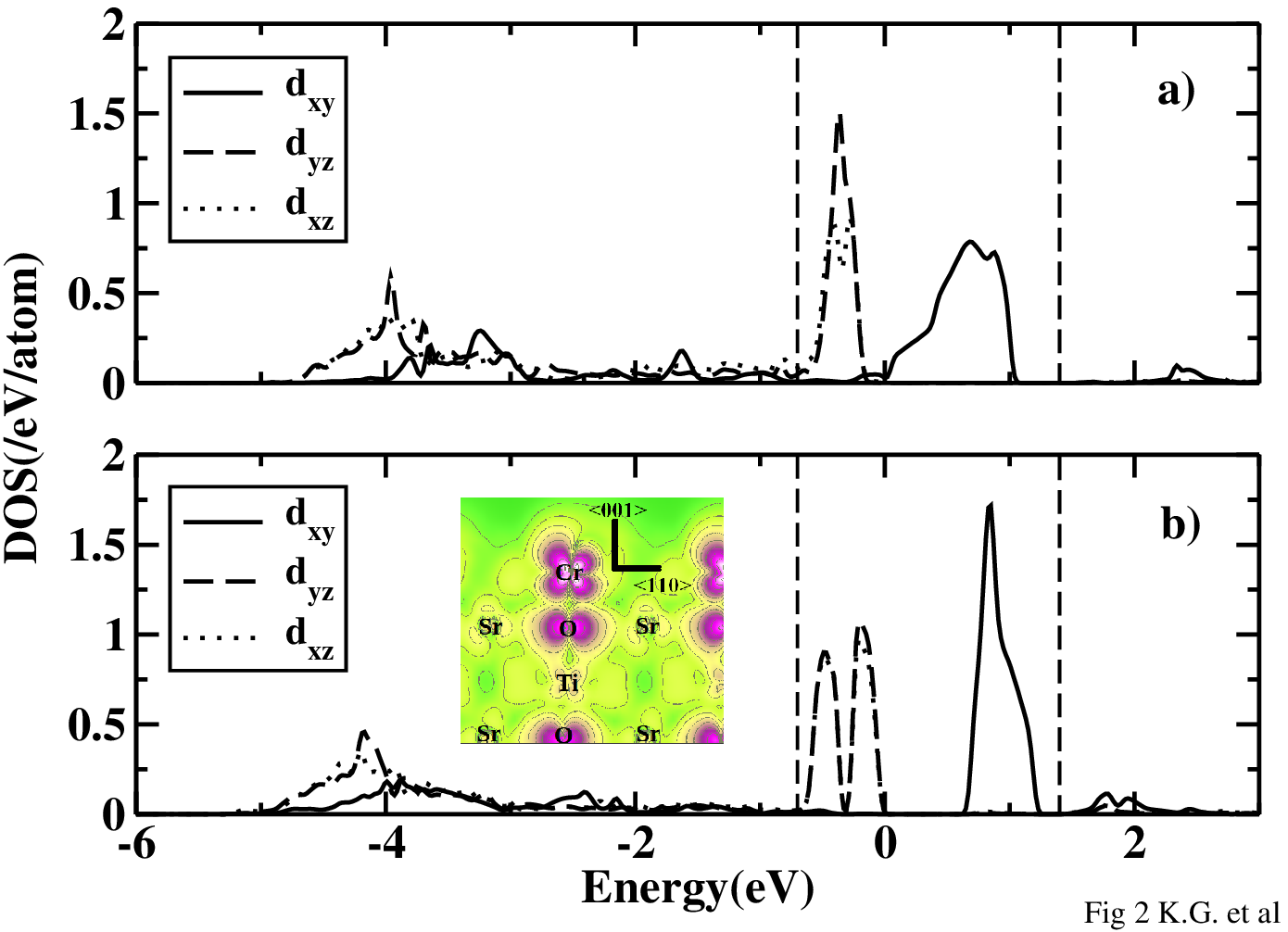}
\caption{ (Color online) The majority spin d$_{xz}$, d$_{yz}$ and d$_{xy}$ projected partial density of states 
for a Cr atom in (a) bulk-like geometry, (b) in the optimised ferroelectric geometry for two 
MLs of SrCrO$_3$ films on SrTiO$_3$. 
The charge density associated with the antibonding states
with $d_{xz}$, $d_{yz}$ character is shown in inset of panel (b).
Vertical dashed lines indicate the valence band maximum and
conduction band minimum of SrTiO$_3$. The zero of energy indicates the fermi energy. }

\end{figure}

\begin{figure}
\includegraphics[scale=0.5,keepaspectratio=true,angle=0]{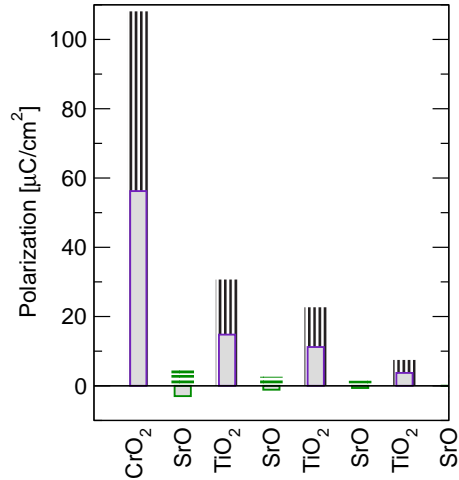}
\caption{(Color online) The layer resolved polarizations for two MLs of SrCrO$_3$ on SrTiO$_3$ : 
The bars with vertical, violet pattern show the total (ionic plus electronic) polarization 
in CrO$_2$ layers, while the ones with the horizontal green pattern show the total polarization 
in SrO layers. The gray bars outlined with violet (CrO$_2$ layers) or green (SrO layers) indicate 
the electronic contribution to the total polarization in the respective layers.}
\end{figure}

\begin{figure}
\includegraphics[scale=0.5,keepaspectratio=true,angle=0]{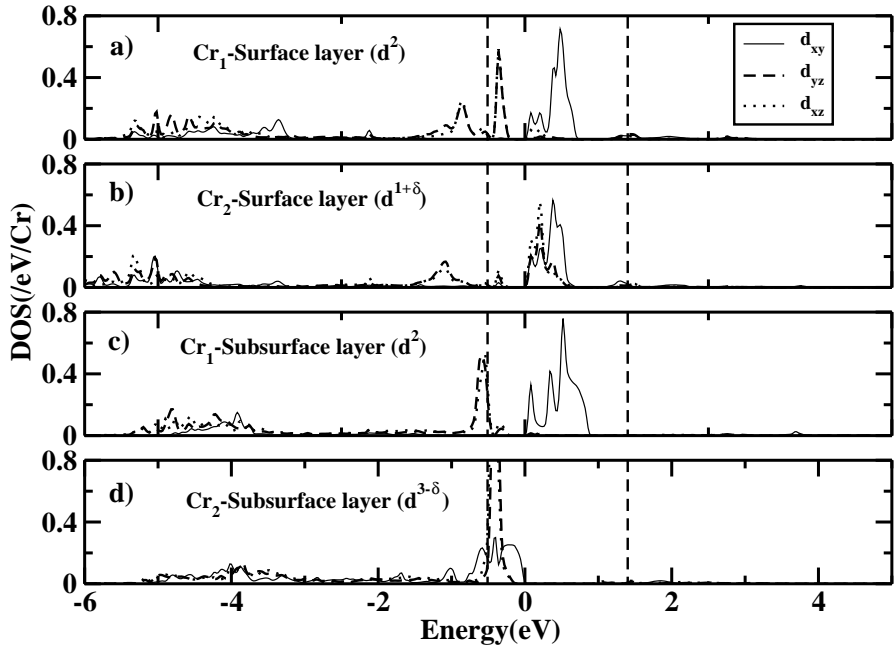}
\caption{ The majority spin d$_{xz}$, d$_{yz}$ and d$_{xy}$ projected partial density of states
for a Cr atom belonging to the surface layer ((a)-(b)), and  to the sub-surface layer (panels (c)-(d)), 
of 4 MLs SrCrO$_3$ films on SrTiO$_3$.
Vertical dashed lines indicate the valence band maximum and
conduction band minimum of SrTiO$_3$. The zero of energy indicates the fermi energy. }

\end{figure}


\begin{references}
\bibitem{details} The ferroelectric simulation cell has an energy of 0.84 eV
above the centrosymmetric one. As all transition metal atoms
show off-centering, dividing by the number of transition metal atoms gives
us a rough idea of the energy for flipping.
\end{references}
\end{document}